\documentclass[a4paper,12pt]{article}

\usepackage{latexsym}

\begin{document}

\title{On the twin paradox in static spacetimes: I. Schwarzschild metric} 
\author{Leszek M. SOKO\L{}OWSKI \\
Astronomical Observatory, Jagiellonian University\\ 
and Copernicus Center for Interdisciplinary Studies,\\
Orla 171,  Krak\'ow 30-244, Poland \\
email: lech.sokolowski@uj.edu.pl} 

\date{}
\maketitle


\begin{abstract}
Motivated by a conjecture put forward by Abramowicz and Bajtlik we reconsider the twin paradox 
 in static spacetimes. According to a well known theorem in Lorentzian geometry the longest 
 timelike worldline between two given points is the unique geodesic line without points 
 conjugate to the initial point on the segment joining the two points. We calculate the proper 
 times for static twins, for twins moving on a circular orbit (if it is a geodesic) around a 
 centre of symmetry and for twins travelling on outgoing and ingoing radial timelike 
 geodesics. We show that the twins on the radial geodesic worldlines are always the oldest 
 ones and we explicitly find the the conjugate points (if they exist) outside the relevant 
 segments. As it is of its own mathematical interest, we find general Jacobi vector fields on the 
 geodesic lines under consideration. In the first part of the work we investigate 
 Schwarzschild geometry.
 PACS number: 04.50.Kd\\
Keywords: twin paradox, static spacetimes, Jacobi fields, conjugate points.\\

PACS{04.20Jb}

\end{abstract}

\section{Introduction}
 The twin paradox is well understood in Minkowski spacetime whereas in curved spacetimes less 
 is known about generic features of the twins following various worldlines and recent works studying 
 various special cases are quite numerous (see e.~g.~\cite{LI}, \cite{Io1}, \cite{Io2}, \cite{JW}, 
 \cite{DG}, \cite{Sz} and references therein).
 Abramowicz, Bajtlik and Klu\'{z}niak investigated a couple of static spacetimes where the 
 absolute rest is defined by the unique timelike Killing vector field and a stationary axially 
 symmetric spacetime where the standard of rest is determined by the rest frame of the zero 
 angular momentum observers (ZAMO) and calculated the proper times for the twins orbiting on 
 the same circular orbit with different constant linear velocities. From the case of these 
 circular and stationary motions they have drawn a conclusion which being generalized to other 
 worldlines is actually a conjecture: 'in all situations in which the absolute motion may be 
 defined in terms of some invariant global properties of the spacetime, the twin who moves 
 faster with respect to the global standard of rest is younger at the reunion, irrespectively 
 to twins' accelerations' \cite{ABK}, \cite{AB}. It is the purpose of the present work to show that in 
 general the conjecture is false.\\
 
 We first make a comment on the concept of rest. Geometrically an absolute standard of rest is 
 provided by the timelike Killing vector field (if it exists and is unique), one may, however, 
 consider a universal standard of rest based on the synchronous (i.e.~comoving, i.e.~normal 
 Gauss) frame of reference which is freely falling in the gravitational field. The comoving 
 frame may cover only a part of the spacetime (as it develops, besides some special cases, 
 coordinate singularities) and the distances between freely falling bodies change in time, 
 nevertheless to some extent the bodies may be treated as being at rest. Accordingly, if 
 radially freely falling bodies in Schwarzschild spacetime are supposed to be at rest, then a 
 static Killing observer and an observer on a circular geodesic (i.e.~an observer orbiting 
 around the central body) are moving with respect to them and, as is shown below, their proper 
 times are shorter than those at rest. This case supports the Abramowicz's, Bajtlik and 
 Klu\'{z}niak (ABK) conjecture. However, this notion of rest is not unambiguous. In fact, the 
 observers on circular orbits are freely falling too and if they are regarded as staying at rest 
 then the radially falling observers and the static Killing ones are in respective motion and 
 their proper times are larger than those of the motionless (i.e.~orbiting on circles) 
 observers. We therefore assume that the state of rest is determined by the unique Killing 
 field. \\
 
 The problem of which twin travelling between two given spacetime points is older is of purely 
 geometrical nature and the answer to the question of which worldline makes a twin the oldest 
 one (assuming a large number of twins following various worldlines with the same endpoints) 
 is implicitly well known as it may be found in advanced monographs. Imprecise answer is even 
 contained in an intermediate level textbook \cite{St}: 'for a really reliable answer one has 
 to know how to deal with accelerated systems; here general relativity is to be asked, and 
 answer is: yes, travelling (deviating from geodesic motion) keeps you younger'.\\
 
 A rigorous formulation and solution of the problem is following. Let $U$ be a convex normal 
 neighbourhood of a point $p$ in any Lorentzian spacetime. Then the following proposition holds:\\
 \textsc{Proposition 1} (proposition 4.5.3 in \cite{HE})\\
 In any convex normal neighbourhood $U$, if $p$ and $q$ can be joined by a timelike curve 
 then the unique timelike geodesic connecting them has length strictly greater than that of 
 any other piecewise smooth timelike curve between the points.
 
 Schwarzschild spacetime is misleading in this context. Consider two twins: A is at rest at a 
 point $r=r_0>3M$ (in the standard Schwarzschild coordinates) and B moves on a timelike geodesic 
 line with a circular orbit in the space at $r=r_0$. Initially the twins are at a spacetime 
 point $P_0$ ($t=t_0$, $r=r_0$), then B flies away, makes a full circle around the centre and 
 meets A again at $P_1$ ($t=t_1$, $r=r_0$). A simple calculation shows that the proper time 
 $s_A$ (i.e.~the worldline length) of A between $P_0$ and $P_1$ is longer than the geodesic 
 proper time $s_B$ of B, 
 \begin{equation}\label{n1}
\frac{s_A}{s_B}=\frac{1}{\sqrt{1-\left(\frac{v}{c}\right)^2}}=\left(\frac{r_0-2M}{r_0-3M}
\right)^{1/2}>1,
\end{equation} 
where $v$ is B's velocity with respect to the static A \cite{ABK}, \cite{AB}. This case gave rise 
to the conclusion that the twin who moves faster (and without acceleration) is younger than the 
static twin (subject to acceleration). However this simple example merely shows that $P_1$ 
does not belong to a convex normal neighbourhood of $P_0$. In fact, there is at least one timelike 
geodesic line joining $P_0$ and $P_1$ whose length is larger than that of B.\\

Here the key notion is that of conjugate points and for the reader's convenience we briefly 
recall the relevant facts. Let $\gamma$ be a timelike geodesic with tangent unit $u^{\alpha}$. 
Any vector field $Z^{\mu}(s)$ which is a solution to the geodesic deviation equation on $\gamma$ 
is called a Jacobi field on $\gamma$. A pair of points $p$ and $q$ on $\gamma$ are said to be 
conjugate if there exists a Jacobi field $Z^{\mu}$ on $\gamma$ which is not identically zero 
and $Z^{\mu}(p)=Z^{\mu}(q)=0$ (\cite{HE} chapter 4). Points $p$ and $q$ are conjugate if 
infinitesimally nearby geodesic intersects $\gamma$ at both $p$ and $q$. Existence of conjugate 
points on a geodesic shows that it is neither the unique geodesic nor the longest curve 
connecting its endpoints. More precisely, if a timelike geodesic $\gamma$ joining points $p_1$ 
and $p_2$ has a point $q$ conjugate to $p_1$ belonging to the segment $p_1p_2$, then there 
exists a nearby timelike curve $\tau$ (not necessarily a geodesic one) with endpoints $p_1$ 
and $p_2$, which is longer than $\gamma$, $s(\tau)>s(\gamma)$. Conversely, if there is no 
conjugate points on a timelike geodesic, then it is the longest curve between its endpoints, 
independently of whether the endpoints lie in some convex normal neighbourhood, or not. It is 
stated in\\
\textsc{Proposition 2} (proposition 4.5.8 in \cite{HE})\\ 
A timelike geodesic $\gamma$ has the maximal length from $p_1$ to $p_2$ if and only if there is 
no point conjugate to $p_1$ on the segment $p_1p_2$.
 In some spacetimes, e.g.~anti--de Sitter space, two given points may be connected by various 
 timelike curves, but none of these attains maximal length. The longest curve always exists 
 in globally hyperbolic spacetimes.\\
 \textsc{Proposition 3} (proposition 9.4.4 in \cite{HE})\\
 Let $(M,g_{\alpha\beta})$ be a globally hyperbolic spacetime. Let $p_1$ and $p_2$ be connected 
 by timelike curves. Then there exists a timelike geodesic $\gamma$ from $p_1$ to $p_2$ 
 having the maximal length.
 The existence of conjugate points is determined by\\
 \textsc{Proposition 4} (proposition 4.4.2 in \cite{HE})\\
 If $R_{\alpha\beta}\,u^{\alpha}\,u^{\beta}\geq 0$ on a timelike geodesic $\gamma$ and if the 
 tidal force $R_{\mu\alpha\nu\beta}\,u^{\alpha}\,u^{\beta}\neq 0$ at some point $p_0$ on 
 $\gamma$, there will be conjugate points $p$ and $q$ on $\gamma$, providing that the geodesic 
 can be extended sufficiently far.
 
 Returning to the problem of the rate of ageing of the twins one sees that neither the 
 velocity (in stationary spacetimes) nor absence of acceleration (on geodesic worldlines) alone 
 is sufficient to establish which twin will be older at the reunion. In a curved spacetime 
 there is a great variety of motions and for a given pair of worldlines there is no generic 
 criterion (if none of them is a geodesic free of conjugate points) stating which of these is 
 longer; one must explicitly compute their lengths. Therefore besides studying case by case 
 one can only determine the longest worldlines joining given two points in some spacetimes. 
 This is the heart of the present work.\\
 
 For a given pair of spacetime points one applies physical arguments to find out a timelike 
 geodesic connecting them and seeks for conjugate points on it. In physically interesting 
 spacetimes the strong energy condition holds implying $R_{\alpha\beta}\,u^{\alpha}\,u^{\beta}\geq 0$ 
 and in most cases the tidal forces do not vanish, hence Proposition 4 indicates that the 
 geodesic under consideration contains somewhere conjugate points. It is then crucial to 
 establish whether the conjugate points belong to the relevant segment of the geodesic.\\
 In this work we explicitly analytically compute location of conjugate points 
 on physically distinguished timelike geodesic lines in a number of static spacetimes and for
 comparison in Robertson--Walker spatially flat universe. To establish that the given 
 geodesic is not the maximal one it is sufficient to find out a particular Jacobi vector field 
 on it which reveals conjugate points within the relevant segment. A more general problem is 
 finding out a generic Jacobi field on the given geodesic curve and there is an opinion 
 (private communication) of a few experts in the field that this work is worth doing it as it 
 is of mathematical interest. We get general solutions of the geodesic deviation equation for 
 worldlines followed by the twins considered in the problem of their ageing rates. 
 In the first part of the work we discuss the most convenient form of the geodesic deviation 
 equation for the Jacobi field and then deal with Schwarzschild spacetime.
 
 \section{Timelike geodesics in static spacetimes and Jacobi fields}
 We consider static spacetimes since there the orbits of the timelike Killing vector field are 
 orthogonal to hypersurfaces of simultaneity and the notion of observers at rest is unambiguous. 
 The case of standard of rest with respect to the ZAMO frames shows that stationary non--static 
 spacetimes require a separate treatment \cite{AB}. In a static spacetime there always exists a 
 coordinate chart in which the metric takes the form 
 \begin{equation}\label{n2}
 ds^2=g_{00}(x^k)\,dt^2+g_{ij}(x^k)\,dx^i\,dx^j
\end{equation} 
 and the timelike Killing field is $K^{\alpha}=\delta^{\alpha}_0$, $K^{\alpha}K_{\alpha}
 >0$\footnote{For obvious reasons we choose the metric signature $(+---)$.}, 
 $\alpha,\beta=0,1,2,3$ and $i,j,k=1,2,3$. An observer is at rest if his worldline coincides 
 with one of the orbits of the Killing field, then his velocity is $u^{\alpha}=g_{00}^{-1/2}\,
 \delta^{\alpha}_0$. From the geodesic equation $(D/ds)u^{\alpha}=0$, where $D/ds$ is the 
 absolute derivative along a curve with respect to its arc length, one immediately sees that a 
 static observer worldline is a geodesic if and only if $g_{00}=1$. This is a very special case 
 and none of the physically interesting static spacetimes satisfies this condition. We shall 
 always denote by A the nongeodesic observer at rest.\\
 
 We assume that the spatial coordinates $(x^i)$ are so chosen that $(\partial/\partial x^1)
 g_{00}>0$. The coordinate $x^1$ varies in the range $-\infty\leq a\leq x^1\leq b\leq+\infty$ 
 and $x^1$ tending to $b$ corresponds to spatial infinity (if it exists). Due to these properties 
 we dub $x^1$ into 'radial coordinate'. We then seek for two kinds of worldlines as possible 
 geodesics. Firstly, 'circular orbits' with $x^1=\textrm{const}\neq 0$ which are spatially 
 closed and secondly 'radial worldlines' with $x^2$ and $x^3$ constant and $x^1$ varying in an 
 subinterval $a_0\leq x^1\leq b_0$ of $(a,b)$. In the spacetimes studied in this 
 work the radial curves are timelike geodesics.\\
 In the physical interpretation we shall assume that the twin B follows a circular worldline 
 and the twin C moves on a radial geodesic curve\footnote{We use here the traditional term 
 'twins' instead of the correct 'triplets'.}. The twin A stays at $x^1=a_0$ and in a  
 coordinate time interval $\Delta t$ his proper time is $s_A=\sqrt{g_{00}}\,\Delta t$. 
 The twin C's radial worldline emanates outwards from $x^1=a_0$, reaches a maximal height 
 $x^1=b_0$, turns inwards and comes back to $x^1=a_0$ to meet again A and possibly B. The 
 corresponding proper time $s_C$ of the twin C may be greater than $s_A$; this is possible 
 although the expression $g_{ij}\,dx^i\,dx^j<0$ for $dx^i\neq 0$ if $g_{00}$ grows sufficiently 
 quickly in the interval $a_0<x^1<b_0$.\\
 
 On a given radial or circular geodesic $\gamma$ the conjugate points can be determined directly 
 from the definition: by finding a generic solution $Z^{\mu}(s)$ to the geodesic deviation equation 
 \begin{equation}\label{n3}
 \frac{D^2}{ds^2}\,Z^{\mu}=R^{\mu}{}_{\alpha\beta\gamma}\,u^{\alpha}\,u^{\beta}\,Z^{\gamma},
\end{equation} 
 where $u^{\alpha}=dx^{\alpha}/ds$, $u^{\alpha}u_{\alpha}=1$, is the tangent vector field to 
 the geodesic. The Jacobi field is orthogonal to the velocity vector, $Z^{\alpha}u_{\alpha}=0$. 
 Due to the presence of the absolute derivatives, equations (3) are very complicated ODEs of 
 second order and to simplify them one applies the following approach (\cite{HE} chapter 4 and 
 \cite{W} chapter 9). On $\gamma$ one introduces an orthonormal triad of spacelike vector 
 fields $e_{a}^{\mu}(s)$, $a=1,2,3$, which are orthogonal to $u^{\mu}$,
 \begin{equation}\label{n4}
 e_{a}^{\mu}\, e_{b\mu}=-\delta_{ab}, \qquad  e_{a}^{\mu}\,u_{\mu}=0.
\end{equation} 
 $Z^{\mu}$ are components of the Jacobi field in the coordinate basis, $\mathbf{Z}=Z^{\mu}
 \partial_{\mu}$. Since $Z^{\alpha}u_{\alpha}=0$, the field may be decomposed in the triad 
 basis, $\mathbf{Z}=\sum_{a} Z_a\,\mathbf{e}_a$ and when the triad fields are kept fixed the 
 functions $Z_a(s)$ are three scalar fields on $\gamma$. Equations (3) get simplified when 
 $e_{a}^{\mu}$ are parallelly transported along $\gamma$, $(D/ds)e_{a}^{\mu}=0$. The components 
 $Z^{\mu}$ are expressed in terms of the scalars as 
 \begin{equation}\label{n5}
 Z^{\mu}=\sum_{a=1}^{3} Z_a\,e_{a}^{\mu}.
\end{equation} 
Inserting (5) into (3) and multiplying the result by $e_{b}^{\mu}$ one arrives at the following 
linear equations for the three scalars $Z_a$,
 \begin{equation}\label{n6}
 \frac{d^2}{ds^2}\,Z_a=-e_{a}^{\mu}\,R_{\mu\alpha\beta\gamma}\,u^{\alpha}\,u^{\beta}
 \sum_{b=1}^{3}e_{b}^{\gamma}\,Z_b.
\end{equation} 
 At an initial point $p_0$ on $\gamma$ one gives initial values $e_{a}^{\mu}(p_0)$ satisfying 
 (4), then the triad fields are uniquely determined by the parallel propagation. One seeks for 
 a general solution to (6); for each $a$ it depends on two integration constants and one of 
 them is a multiplicative factor. In the search for conjugate points one then imposes the initial 
 condition $Z_a(p_0)=0$ and gets the desired Jacobi field. If one of the scalars, say $Z_1$, 
 vanishes at a point $q$ on $\gamma$, $Z_1(q)=0$, one can identically set $Z_2=Z_3\equiv 0$ and 
 the Jacobi field $Z^{\mu}(s)=Z_1(s)\,e_{1}^{\mu}(s)$ is zero at $q$; the points $p_0$ and $q$ 
 are conjugate on $\gamma$.\\
 
 In the next section we calculate and compare the lengths of a circular and a radial geodesic 
 worldlines, determine Jacobi fields on them and their conjugate points in the case of the 
 static spherically symmetric vacuum spacetime.
 
 \section{Schwarzschild spacetime}
 In this spacetime (in standard coordinates $t,r,\theta,\phi$) geodesic circular orbits exist only 
 for $r=r_0>3M$, where $M$ is the energy of the spacetime\footnote{Usually we put $c=G=1$; from 
 time to time we shall explicitly display $c$ for clarity.}. All the motions take place in the 
 two--surface $\theta=\pi/2$.\\
 The twin A stays at rest at $r=r_0$, $\theta=\pi/2$ and $\phi=\phi_0$. The twin B moves on a geodesic 
 circular orbit $r=r_0$ with a constant linear velocity $v_B$ with respect to A. All the 
 parameters of the B's geodesic are determined by $r_0$ (and $M$) and it is parameterized by its 
 length, 
 \begin{equation}\label{n7}
 t-t_0=\left(\frac{r_0}{r_0-3M}\right)^{1/2}\,s, \qquad 
 \phi-\phi_0=\frac{1}{r_0}\left(\frac{r_0}{r_0-3M}\right)^{1/2}\,s
\end{equation} 
 and $(v_B/c)^2=M/(r_0-2M)$; $t_0$ is the moment when A, B and C are together at the same place. 
 The length of the geodesic B after making one full circle is determined from (7) as $\phi-\phi_0=
 2\pi$ and is 
 \begin{equation}\label{n8}
 s_B=2\pi r_0 \left(\frac{r_0-3M}{M}\right)^{1/2}
\end{equation} 
and the corresponding coordinate time interval is 
 \begin{equation}\label{n9}
 \Delta t=2\pi r_0\left(\frac{r_0}{M}\right)^{1/2}.
\end{equation}
 By assumption at the point $P_0(t=t_0, r=r_0, \phi=\phi_0)$ the worldlines of the three twins 
 intersect. The twin C moves on a radial geodesic emanating from $P_0$ with an initial radial velocity 
 $(dr/ds)(t_0)=u>0$ directed outwards at $\theta=\pi/2$ and $\phi=\phi_0$. At $t=t_M$ he reaches a 
 maximum height $r=r_M$ where his radial velocity decreases to zero and he falls down along the 
 ingoing radial geodesic until at $t=t_1$ he returns to $r=r_0$. Since the spacetime is static, the 
 coordinate and proper time of the flight upwards are equal to these of falling down, i.e. 
 $t_1-t_M=t_M-t_0$.\\
 By assumption the three twins meet again at $P_1(t=t_1, r=r_0, \phi=\phi_0)$, i.e.~C falls back 
 to the initial place when B makes a full circle,
\begin{equation}\label{n10}
 t_1-t_0=2(t_M-t_0)=\Delta t.
\end{equation} 
 The proper time $s_A$ of the static twin from $P_0$ to $P_1$ is 
 \begin{equation}\label{n11}
 s_A= \left(1-\frac{2M}{r_0}\right)^{1/2}\,\Delta t=2\pi r_0\left(\frac{r_0}{M}-2\right)^{1/2}
\end{equation}  
 and according to (1) is larger than $s_B$ \cite{ABK}, \cite{AB}, what shows that the circular 
 geodesic B contains conjugate points between $P_0$  and $P_1$.\\
 One then conjectures that the longest curve connecting $P_0$ and $P_1$ is the radial geodesic C. 
 To confirm the conjecture one computes the C's length and then shows that conjugate points, if 
 exist on C, are located beyond the segment $P_0P_1$. As is well known it is convenient to 
 parameterize a radial geodesic in this spacetime by an angular parameter $\eta$,
 \begin{equation}\label{n12}
 r(\eta)=r_M\,\cos^2\eta=\frac{1}{2}r_M(\cos2\eta+1),
\end{equation}  
 where $-\alpha/2\leq\eta<\pi/2$ and $r_M$ is the largest radial distance from the centre. By 
 definition the initial point is $r(-\alpha/2)=r_0$, then 
 \begin{equation}\label{n13}
\cos\alpha=\frac{2r_0}{r_M}-1 \qquad \textrm{and} \qquad 0<\alpha<\pi.
\end{equation}  
 The return point is $r(\alpha/2)=r_0$, the event horizon $r=2M$ is at $\eta=\eta_H$ where 
 $\cos^2\eta_H=2M/r_M$ and $r(\pi/2)=0$. On the radial geodesic all geometrical quantities are 
 determined by $r_0$ and the initial velocity $u$, or equivalently by the integral of energy. The 
 twin C with mass $m$ and 4--momentum $p^{\alpha}$ has conserved energy $E$ determined by the 
 timelike Killing field, which is normalized to unity at the spatial infinity, $K^{\alpha}=
 \delta^{\alpha}_0$, hence\\
 $E/c=K^{\alpha}\,p_{\alpha}=p_0=(1-2M/(c^2r))mc\,dx^0/ds>0$.\\
 We introduce dimensionless constant energy per unit rest energy, $k\equiv E/(mc^2)$, then 
 \begin{equation}\label{n14}
\frac{dt}{ds}=k\left(1-\frac{2M}{r}\right)^{-1},
\end{equation}  
where $x^0\equiv t$. The geodesic equation for the radial coordinate is replaced by the universal 
integral of motion $u^{\alpha}\,u_{\alpha}=1$ which in the present case reads
 \begin{equation}\label{n15}
\left(\frac{dr}{ds}\right)^2=k^2+\frac{2M}{r}-1.
\end{equation} 
 At the starting point one has 
 \begin{equation}\label{n16}
u^2=k^2+\frac{2M}{r_0}-1
\end{equation} 
 and $r_M$ is determined from $k^2-1+\frac{2M}{r_M}=0$, hence
 \begin{equation}\label{n17}
r_M=\frac{2M}{1-k^2} \qquad \textrm{and} \qquad 0<k<1.
\end{equation} 
Since stable circular orbits require $r_0>3M$ and $r_M>r_0$ one gets a lower limit 
$k^2>1/3$. The tangent vector $u^{\alpha}$ has, from (14) and (15), the nonvanishing components 
 \begin{equation}\label{n18}
\frac{dt}{ds}=\frac{k\cos^2\eta}{\cos^2\eta-1+k^2}, \qquad \frac{dr}{ds}=-\sqrt{1-k^2}\,
\tan\eta;
\end{equation} 
 the minus sign ensures that for $-\alpha/2<\eta<0$ the velocity is directed outwards and for 
 $\eta>0$ it is directed inwards. To obtain relations $t=t(\eta)$ and $s=s(\eta)$, one replaces 
 (18) by derivatives with respect to $\eta$. The derivative $ds/d\eta$ is derived from the 
 formula for $ds^2$ expressed as a function of $\eta$; one gets 
 \begin{equation}\label{n19}
 \left(\frac{ds}{d\eta}\right)^2=\frac{2r_M^3}{M}\,\cos^4\eta
\end{equation} 
 and employing that $ds/d\eta$ is always positive one finds 
 \begin{equation}\label{n20}
s(\eta)=\left(\frac{r_M^3}{2M}\right)^{1/2}\,\left[\frac{1}{2}(\sin2\eta+\sin\alpha)+\eta+
\frac{\alpha}{2}\right].
\end{equation} 
 This immediately yields the length of C from $P_0$ to $P_1$,
 \begin{equation}\label{n21}
s_C=s\left(\frac{\alpha}{2}\right)=\left(\frac{r_M^3}{2M}\right)^{1/2}\,\left[\frac{2r_0}{r_M}
\sqrt{\frac{r_M}{r_0}-1}+\arccos\left(\frac{2r_0}{r_M}-1\right)\right].
\end{equation}
As concerns $t(\eta)$ one has $dt/d\eta=(dt/ds)(ds/d\eta)$ and substituting from (18) and (19) 
one gets 
 \begin{eqnarray}\label{n22}
t(\eta)-t_0 & = & \int^{\eta}_{-\frac{\alpha}{2}} k\,\left(\frac{r_M^3}{2M}\right)^{1/2}\,
\frac{(\cos2\eta+1)^2}{\cos2\eta+2k^2-1}\,d\eta=
\nonumber\\
& = &  k\,\left(\frac{r_M^3}{2M}\right)^{1/2}\,[(3-2k^2)(\eta+\frac{\alpha}{2})+\frac{1}{2}
(\sin2\eta+\sin\alpha)]+
\nonumber\\
& + & 2M\,\ln\left(\frac{b+\tan\eta}{b-\tan\eta}\right)-
2M\,\ln\left(\frac{b-\tan\alpha/2}{b+\tan\alpha/2}\right),
\end{eqnarray} 
 where\footnote{Clearly formulae (20) and (22) have already been known in various versions, we 
 rederive them here for the reader's convenience.} $b=k(1-k^2)^{-1/2}$. Notice that at the 
 event horizon $\sin\eta_H=+k$, then $\tan\eta_H=b$ and $t\rightarrow+\infty$ as it should be. 
 Accordingly, the coordinate time interval of the flight back and forth is 
  \begin{equation}\label{n23}
t(\alpha/2)-t_0 = k\,\left(\frac{r_M^3}{2M}\right)^{1/2}\,
[(3-2k^2)\alpha+\sin\alpha]+4M\,\ln\left(\frac{b+(\frac{r_M}{r_0}-1)^{1/2}}
{b-(\frac{r_M}{r_0}-1)^{1/2}}\right).
\end{equation}
 Here
 \begin{displaymath}
 b^2-\left(\frac{r_M}{r_0}-1\right)=\frac{1}{1-k^2}\,\left(1-\frac{2M}{r_0}\right)>0,
 \end{displaymath}
 hence the denominator of the fraction under the logarithm is always positive. The interval (23) 
 must be equal to $\Delta t$ given in (9) and this gives a transcendental algebraic equation for 
 the energy $k$ or equivalently the initial radial velocity $u$ in terms of the parameter $r_0$.  
 Introducing for convenience a parameter $R=r_0/(2M)>3/2$ and replacing the 
 unknown $k(r_0)$ by $x=r_0/r_M=R\,(1-k^2)=1-u^2\,R$ one finally gets an equation for $x(r_0)$ 
 (or $x(R)$),
\begin{eqnarray}\label{n24}
2\sqrt{2}\pi R^{3/2} & = & \left(1-\frac{x}{R}\right)^{1/2}\left(\frac{R}{x}\right)^{3/2}\,
\left[\left(1+\frac{2x}{R}\right)\alpha+2\sqrt{x-x^2}\right]+
\nonumber\\
& + & 2\ln\left(\frac{\sqrt{R-x}+\sqrt{1-x}}{\sqrt{R-x}-\sqrt{1-x}}\right),
\end{eqnarray}
 where, according to (13), $\cos\alpha=2x-1$. It may be shown that for each value of $R>3/2$ there is  
 only one solution of the equation in the interval $0<x<1$. Inserting this value of $x(R)$ into (21)  
 one gets the ratio of the twins' proper times, 
\begin{equation}\label{n25}
\frac{s_C}{s_A}=\frac{1}{\sqrt{2}\pi}\,\left(1-\frac{1}{R}\right)^{-1/2}\,x^{-3/2}\,
\left[\sqrt{x-x^2}+\frac{1}{2}\arccos(2x-1)\right].
\end{equation}
 Equation (24) was numerically solved for five values of $R$ consecutively growing by 0,1 from 
 1,5 to 2 and for all integers from $R=2$ to 100. The value of $x(R)$ is a monotonic slowly 
 diminishing function of $R$ (we did not attempt to fit it by an analytic expression) and decreases  
 from 0,545246 for $R=1,5$ to 0,44932 for $R=100$. It is hard to analytically establish from eq. (25)  
 which proper time is larger and we compute their 
 numerical values. In table 1 we give the ratio $s_C/s_A$ for some chosen values of $x(R)$. 
 \begin{table}
\caption{The ratio $s_C/s_A$ for 5 values of $x(R)$.}
\begin{tabular}{lll}
\hline\noalign{\smallskip}
 $R$  &  $x$ &   $s_C/s_A$ \\
 \noalign{\smallskip}\hline\noalign{\smallskip}
 1,5  & 0,545246 & 1,19878 \\
 1,6  & 0,534476 & 1,17554 \\
 2    & 0,508559 & 1,12061 \\
 4    & 0,472995 & 1,04849 \\
 100  & 0,44932  & 1,00162 \\
 \noalign{\smallskip}\hline 
\end{tabular}
\end{table}
 As it should be expected, the farther the point $r_0$ is from the event horizon the smaller 
 the ratio $s_C/s_A$ is and always the radial geodesic C is longer than the non--geodesic 
 curve A. That $s_C>s_A$ was previously found in a special case in \cite{GB}. 

 Both the geodesic curves B and C satisfy the conditions of Proposition 4, hence there do exist 
 conjugate points and to identify them we now find Jacobi fields on these worldlines. 
  
 \subsection{The timelike radial geodesic C}
 An alternative expression to formula (18)  for the tangent vector to C in terms of the coordinate 
 $r$ is 
\begin{equation}\label{n26}
u^{\alpha} =\left[\frac{kr}{r-2M},\,\varepsilon\left(\frac{2M}{r}-1+k^2\right)^{1/2}, 0, 0
\right]
\end{equation}
 where $\varepsilon=+1$ on the outwards directed segment of C and $\varepsilon=-1$ on the inwards 
 directed piece. The parallelly propagated and orthogonal to C orthonormal spacelike triad is 
 chosen as 
\begin{eqnarray}\label{n27}
e_1^{\mu} & = & \left[\frac{\varepsilon r}{r-2M}\,\left(\frac{2M}{r}-1+k^2\right)^{1/2}, k, 0, 0
\right]=
\nonumber\\
& = & \left[-(1-k^2)^{1/2}\,\frac{\sin\eta\cos\eta}{\cos^2\eta-1+k^2}, k, 0, 0\right],
\end{eqnarray} 
\begin{equation}\label{n28}
e_2^{\mu}=\left[0, 0,\frac{1}{r}, 0\right], \qquad e_3^{\mu}=\left[0, 0, 0, \frac{1}{r}\right]
\end{equation}
with $\varepsilon$ as in (26). The geodesic deviation equation for the scalars $Z_a$ reduces 
 in the present case to
\begin{equation}\label{n29}
\frac{d^2}{ds^2}Z_1=\frac{2M}{r^3}\,Z_1, 
\end{equation}
\begin{equation}\label{n30}
\frac{d^2}{ds^2}Z_2=-\frac{M}{r^3}\,Z_2, 
\end{equation}
\begin{equation}\label{n31}
\frac{d^2}{ds^2}Z_3=-\frac{M}{r^3}\,Z_3. 
\end{equation}
 Simplicity of these equations is deceptive since their left hand sides contain derivatives with 
 respect to the arc length instead of $r$ or $\eta$. One replaces these second 
 derivatives by derivatives with respect to $\eta$,
\begin{equation}\label{n32}
\frac{dZ_a}{ds}=\frac{p}{\cos2\eta+1}\,\frac{dZ_a}{d\eta},
\end{equation} 
\begin{equation}\label{n33}
\frac{d^2Z_a}{ds^2}=\frac{p^2}{(\cos2\eta+1)^2}\,\left[\frac{d^2Z_a}{d\eta^2}+
\frac{2\sin2\eta}{\cos2\eta+1}\,\frac{dZ_a}{d\eta}\right], 
\end{equation}
 where $p\equiv(r_M^3/2M)^{-1/2}$. Then the Jacobi equations read 
\begin{equation}\label{n34}
\frac{d^2Z_1}{d\eta^2}+\frac{2\sin2\eta}{\cos2\eta+1}\,\frac{dZ_1}{d\eta}-
\frac{8}{\cos2\eta+1}\,Z_1=0, 
\end{equation} 
\begin{equation}\label{n35}
\frac{d^2Z_2}{d\eta^2}+\frac{2\sin2\eta}{\cos2\eta+1}\,\frac{dZ_2}{d\eta}+
\frac{4}{\cos2\eta+1}\,Z_2=0 
\end{equation}
 and for $Z_3$ one has an equation identical to that for $Z_2$. To solve eq. (34) one introduces 
 a change of the independent variable, $x=\cos2\eta$ (not to be confused with the parameter 
 $r_0/r_M$ introduced earlier), then $u=(x+1)/2$ and substitutes 
 $Z_1(u)=4u^2\,G(u)$. Then (34) reduces to a special case of Gauss hypergeometric equation 
 which is solved in elementary functions. After some manipulations one gets a generic solution 
 for $Z_1$, 
\begin{equation}\label{n36}
Z_1(\eta)=C_1\tan\eta+C_2(3\eta\tan\eta-\frac{3}{2}\pi\,\tan\eta-\cos^2\eta+3).
\end{equation}
 In the case of eq. (35) one performs the same operations as previously and finally substitutes 
 $G(u)=\frac{1}{u}\,Y(u)$, this yields an equation 
\begin{equation}\label{n37}
u(u-1)\,\frac{d^2Y}{du^2}+\left(2u-\frac{3}{2}\right)\,\frac{dY}{du}=0,
\end{equation} 
 which is also solved in elementary functions. The general solution is 
\begin{equation}\label{n38}
Z_2(\eta)=C'\sin2\eta+C''(\cos2\eta+1)
\end{equation}  
 and an analogous formula holds for $Z_3$. A general Jacobi field on a radial geodesic was 
 earlier found by Fuchs \cite{F}. He used a different basis triad whose explicit form in terms 
 of $r$ or $\eta$ was not given, hence it is not easy to compare his solutions to ours.\\
 
 We now seek for conjugate points on C. The special Jacobi scalars which vanish at $P_0(\eta=
 -\alpha/2)$ are then 
\begin{equation}\label{n39}
Z_1(\eta)=C_1\left[\left(\frac{3}{2}\alpha+\frac{4\cos\alpha-\cos^2\alpha+5}{2\sin\alpha}+
3\eta\right)\,\tan\eta-\cos^2\eta+3\right],
\end{equation}  
\begin{equation}\label{n40}
Z_2(\eta)=C_2\left[\sin2\eta+\frac{\sin\alpha}{\cos\alpha+1}\left(\cos2\eta+1
\right)\right]
\end{equation}
and $Z_3$ is given by the same formula as that for $Z_2$ with $C_2$ replaced by $C_3$; $C_1, C_2$ 
and $C_3$ are arbitrary constants. Functions $Z_2$ and $Z_3$ vanish at $\eta=\pi/2$, i.e.~at 
$r=0$. In this sense the curvature singularity is a point conjugate on the radial geodesic to 
any point $P_0$. For $-\alpha/2\leq\eta<\pi/2$ the equation $Z_2(\eta)=0$ is reduced to 
\begin{displaymath}
\tan\eta=-\frac{\sin\alpha}{\cos\alpha+1}=-\sqrt{\frac{r_M}{r_0}-1}<0
\end{displaymath} 
and its solution belongs to the interval $-\pi/2<\eta<0$ where tangens is monotonic, hence the 
only solution is $\eta=-\alpha/2$. In sum, the Jacobi field $Z^{\mu}=Z_2e_2^{\mu}+Z_3
e_3^{\mu}$ is different from zero on the radial geodesic segment from any $P_0$ to the event 
horizon.\\
 As concerns $Z_1(\eta)$ one easily shows that the ratio $Z_1/C_1\geq 2$ in the interval 
 $0\leq\eta<\pi/2$ and possible zeros may be only on the outgoing segment of the geodesic, 
 $-\alpha/2<\eta<0$. One views $Z_1$ as a function of two variables, $Z_1(\eta,\alpha)$ with 
 $\alpha\in(0,\pi)$. For any fixed value of $\eta$ one has $\alpha\in(-2\eta,\pi)$ and in 
 this interval $Z_1(\eta,\alpha)$ grows since 
\begin{displaymath}
\frac{\partial Z_1}{\partial\alpha}=\cot^2\alpha/2\,\cos^2\alpha/2\,\tan(-\eta)>0.
\end{displaymath}  
 From $Z_1(-\alpha/2,\alpha)=0$ one infers that $Z_1/C_1$ is positive for $-\alpha/2<\eta<0$, 
 thence the function has no zeros for $-\alpha/2<\eta<\pi/2$.\\
 
 This completes the proof that a timelike radial geodesic consisting of an outgoing segment 
 and an ingoing one contains no conjugate points and is the longest curve joining points $P_0$ 
 and $P_1$ outside the event horizon. Clearly the same statement concerning non-existence of 
 conjugate points applies to any timelike radial geodesic which is purely outgoing or 
 ingoing.
 
 \subsection{The timelike circular geodesic B}
 It is rather surprising that solving the equations for Jacobi fields is in this case much 
 harder than for a radial geodesic. One immediately gets the vector $u^{\alpha}$ tangent to the 
 geodesic B from its parametric form (7). A triad of spacelike orthonormal vector vector fields 
 on B satisfying (4) and parallelly transported along it may be chosen as 
\begin{eqnarray}\label{n41}
e_1^{\mu} & = &  \left[-\left(\frac{Mr_0}{(r_0-2M)(r_0-3M)}\right)^{1/2}\,\sin qs, 
 \left(\frac{r_0-2M}{r_0}\right)^{1/2}\,\cos qs,  0,
\right.
\nonumber\\
& & \left.
{}\frac{-1}{r_0}\,\left(\frac{r_0-2M}{r_0-3M}\right)^{1/2}\,\sin qs\right],
\end{eqnarray} 
\begin{equation}\label{n42}
e_2^{\mu} =\left[0, 0, \frac{1}{r_0}, 0\right],
\end{equation}
\begin{eqnarray}\label{n43}
e_3^{\mu} & = & \left[\left(\frac{Mr_0}{(r_0-2M)(r_0-3M)}\right)^{1/2}\,\cos qs, 
\left(\frac{r_0-2M}{r_0}\right)^{1/2}\,\sin qs, 0,
\right.
\nonumber\\
& & \left.
{}\frac{1}{r_0}\,\left(\frac{r_0-2M}{r_0-3M}\right)^{1/2}\,\cos qs\right],
\end{eqnarray} 
 where $q^2=M/r_0^3$. Employing 
\begin{equation}\label{n44}
u^{\alpha}=(r_0-3M)^{-1/2}\,\left(r_0^{1/2}\delta^{\alpha}_0+\frac{M^{1/2}}{r_0}
\delta^{\alpha}_3\right)
\end{equation} 
 and denoting 
\begin{displaymath}
\beta=3\,\frac{r_0-2M}{r_0-3M}
\end{displaymath}   
 one arrives at the following form of eqs. (6) for the scalars $Z_a(s)$,
\begin{equation}\label{n45}
\frac{d^2}{ds^2}Z_1=q^2\,[(\beta\cos^2qs-1)\,Z_1+\beta\,Z_3\,\sin qs \cos qs],
\end{equation}
\begin{equation}\label{n46}
\frac{d^2}{ds^2}Z_2=-\frac{M}{r_0^2(r_0-3M)}\,Z_2, 
\end{equation}
\begin{equation}\label{n47}
\frac{d^2}{ds^2}Z_3=q^2\,[\beta\,Z_1\,\sin qs \cos qs+(\beta\sin^2qs-1)\,Z_3]. 
\end{equation}

 It will be shown that Jacobi fields spanned on the basis vectors $e_1^{\mu}$ and 
 $e_3^{\mu}$ do not give rise to a point conjugate to $P_0$ on the segment $P_0P_1$ and in 
 this sense they are irrelevant to the twin paradox. We therefore postpone solving equations 
 (45) and (47) to the Appendix and in the current subsection we discuss consequences of the 
 equation for $Z_2$.\\
 
 The general solution for eq.~(46) reads
\begin{equation}\label{n48}
Z_2=C'\,\sin p_0s+C''\,\cos p_0s 
\end{equation}  
where 
\begin{equation}\label{n49}
p_0^2=\frac{M}{r_0^2(r_0-3M)} 
\end{equation}
and $C'$ and $C''$ are integration constants. The special Jacobi field proportional to $e_2^{\mu}$ 
and vanishing at $P_0 (s=0)$ is simply $Z^{\mu}=C\,\delta_2^{\mu}\,\sin p_0s$. The field generates  
 infinite number of points conjugate to $P_0$ which are located at 
\begin{equation}\label{n50}
s_n=n\pi\,
\left(\frac{r_0-3M}{M}\right)^{1/2}\,r_0,
\end{equation}
$n=1,2,\ldots$. The nearest to $P_0$ conjugate point $Q$ is at $s_Q=s_B/2$, that is half way 
between $P_0$ and the endpoint $P_1$. Notice that $P_1$ is also conjugate to both $P_0$ and to 
$Q$.\\

One conjectures that analogous features occur in Reissner--Nordstr\"{o}m spacetime, this will be 
 shown in a forthcoming paper.

 \section{Summary}
In this work we give explicit forms of generic Jacobi fields and conjugate points generated by 
these fields on physically distinguished timelike geodesic curves in Schwarzschild spacetime. 
These are circular orbits and radial curves going back and forth. Applying the theorem that 
the longest timelike curve joining two given points is a geodesic free of points conjugate to its 
endpoints and lying between the ends, we show that in this spacetime the longest worldline 
is any radial geodesic. This outcome may be physically interpreted in terms of the famous 
twin paradox in the spacetime. Considering a number of twins following different worldlines 
with common endpoints one finds that contrary to a recent conjecture the relative ageing 
of the twins is determined neither by their velocities (with respect to a static observer) 
nor their accelerations (in the case of non-geodesic motions). The twin travelling on a 
radial geodesic in Schwarzschild spacetime is at the reunion the oldest one among all the 
twins. Yet the twin moving on a geodesic circular orbit is younger than the static twin.\\

\textbf{Acknowledgements}.
I am deeply indebted to Zdzis\l{}aw Golda for solving a differential equation with the aid of 
Mathematica and to Sebastian Szybka for making some analytic and numerical computations. This 
work was supported by a grant from the John Templeton Foundation.

 \section*{Appendix}
 \renewcommand{\theequation}{\Alph{section}.\arabic{equation}}
 \setcounter{section}{1}
 \setcounter{equation}{0}
 Here we solve equations (45) and (47) and get a generic Jacobi field on the timelike circular 
 geodesic B what allows one for the search for all conjugate points on this geodesic. 
To solve for $Z_1$ and $Z_3$ one first introduces 
 a dimensionless independent variable $x=qs$ ($q$ is defined after eq. (43)) and then replaces the 
 two coupled equations (45) and (47) by one equation for $Z_1$ by eliminating $Z_3$. As a result one  
 gets a fourth order equation 
\begin{eqnarray}\label{nA.1}
Z_1^{(IV)} & - & 4\cot 2x\,Z_1^{'''}+\left(\frac{8}{\sin^22x}-\beta-2\right)\,Z_1^{''}+
\frac{2}{\sin2x}[\beta+(\beta-2)\cos2x]\,Z_1{'} 
\nonumber\\ 
& - & \left[\frac{4}{\sin^22x}\,(\beta\cos2x+
\beta-2)+3-\beta\right]\,Z_1=0 
\end{eqnarray}
with $Z_1^{'}=dZ_1/dx$ etc. The equation has four linearly independent special 
solutions\footnote{The four solutions were found by Z. Golda by manipulating with the 
equation with the aid of Mathematica. There is a first integral generated by the timelike 
Killing vector field (we discuss such integrals in Part II) but it turns out to be of 
little use.} $Z_{1N}$, $N=1,2,3,4$ and these read
\begin{equation}\label{nA.2}
Z_{11}=\sin x,
\end{equation} 
\begin{equation}\label{nA.3}
Z_{12}=2\cos x+\beta x\sin x,
\end{equation}
\begin{equation}\label{nA.4}
Z_{13}=\left\{ \begin{array}{ll}
2\sin x\,\sin(\sqrt{4-\beta}x)+\sqrt{4-\beta}\,\cos x\,\cos(\sqrt{4-\beta}x), & 3<\beta<4,\\
x\cos x+x^2\sin x, & \beta=4,\\
2\sin x\,\sinh(\sqrt{\beta-4}x)+\sqrt{\beta-4}\,\cos x\,\cosh(\sqrt{\beta-4}x), & \beta>4,
\end{array} \right.
\end{equation}
\begin{equation}\label{nA.5}
Z_{14}=\left\{ \begin{array}{ll}
2\sin x\,\cos(\sqrt{4-\beta}x)-\sqrt{4-\beta}\,\cos x\,\sin(\sqrt{4-\beta}x), & 3<\beta<4,\\
4x^3\sin x+(3+6x^2)\,\cos x, & \beta=4,\\
2\sin x\,\cosh(\sqrt{\beta-4}x)+\sqrt{\beta-4}\,\cos x\,\sinh(\sqrt{\beta-4}x), & \beta>4,
\end{array} \right.
\end{equation}
$x=qs$. One sees that $\beta=4$, corresponding to $r_0=6M$, the innermost stable circular orbit 
 (ISCO), is distinguished. The general solution to (A.1) is then 
\begin{displaymath}
Z_1=\sum_{N=1}^{4} C_N\,Z_{1N}
\end{displaymath}
with arbitrary constants $C_N$. The scalar $Z_3$ is determined from eq.~(45) as action of a linear 
operator on $Z_1$,
\begin{equation}\label{nA.6}
Z_3=\left(\frac{2}{\beta\sin2x}\,\frac{d^2}{dx^2}-\cot2x-\frac{\beta-2}{\beta\sin2x}\right)\,
Z_1\equiv LZ_1,
\end{equation}
\begin{equation}\label{nA.7}
\textrm{or} \qquad Z_3=\sum_{N=1} C_N\,LZ_{1N}\equiv\sum_{N=1} C_N\,Z_{3N}. 
\end{equation}
Explicitly, the special solutions $Z_{3N}$ are
\begin{equation}\label{nA.8}
Z_{31}=-\cos x,
\end{equation} 
\begin{equation}\label{nA.9}
Z_{32}=2\sin x-\beta x\cos x,
\end{equation}
\begin{equation}\label{nA.10}
Z_{33}=\left\{ \begin{array}{ll}
-2\cos x\,\sin(\sqrt{4-\beta}x)+\sqrt{4-\beta}\,\sin x\,\cos(\sqrt{4-\beta}x), & 3<\beta<4,\\
x\sin x-x^2\cos x, & \beta=4,\\
\sqrt{\beta-4}\,\sin x\,\cosh(\sqrt{\beta-4}x)-2\cos x\,\sinh(\sqrt{\beta-4}x), & \beta>4,
\end{array} \right.
\end{equation}
\begin{equation}\label{nA.11}
Z_{34}=\left\{ \begin{array}{ll}
-2\cos x\,\cos(\sqrt{4-\beta}x)-\sqrt{4-\beta}\,\sin x\,\sin(\sqrt{4-\beta}x), & 3<\beta<4,\\
3(1+2x^2)\,\sin x-4x^3\,\cos x, & \beta=4,\\
\sqrt{\beta-4}\,\sin x\,\sinh(\sqrt{\beta-4}x)-2\cos x\,\cosh(\sqrt{\beta-4}x), & \beta>4.
\end{array} \right.
\end{equation}
Altogether the general Jacobi field on B depends on 6 arbitrary constants and takes the form 
\begin{equation}\label{nA.12}
Z^{\mu}(s)=\sum_{N=1}^{4} C_N\,(Z_{1N}\,e_1^{\mu}+Z_{3N}\,e_3^{\mu})+(C'\sin p_0s+
C''\cos p_0s)\,e_2^{\mu}. 
\end{equation}

 We have seen in section 3.2 that the special Jacobi field proportional to $e_2^{\mu}$ 
 generates an infinite sequence of points conjugate to $P_0$ of which the first two are 
 located on the segment $P_0P_1$, hence the field does account for the fact that the circular 
 geodesic B is shorter between $P_0$ and $P_1$ than the radial geodesic C and the non-geodesic 
 curve A. We now establish whether a Jacobi field 
spanned on the basis vectors $e_1^{\mu}$ and $e_3^{\mu}$ may also give rise to a conjugate point. 
The procedure is simple and laborious. One must study separate cases $3<\beta<4$, $\beta=4$ and 
$\beta>4$. In the special case $\beta=4$ i.e.~the ISCO geodesic line, the initial condition 
$Z_1(0)=0=Z_3(0)$, yields $C_2=3C_1/4$ and $C_4=-C_1/2$ with arbitrary $C_1$ and $C_3$. It 
turns out that the scalar fields $Z_1(qs)$ and $Z_3(qs)$ have no common roots for $s>0$. In the 
generic case, $\beta\neq 4$, one again gets that the functions $Z_1$ and $Z_3$ vanishing at 
$s=0$ depend on arbitrary $C_1$ and $C_3$, now with 
\begin{equation}\label{nA.13}
C_2=-\frac{1}{2}\,|\beta-4|^{1/2}\,C_3, \qquad C_4=-\frac{1}{2}\,C_1.
\end{equation}
For $\beta>4$, i.e.~for $3M<r_0<6M$, one again finds that $Z_1$ and $Z_3$ do not possess 
common roots for $s>0$. Finally, for $r_0>6M$ one finds that the scalars 
\begin{equation}\label{nA.14}
Z_1=2Z_{11}-Z_{14} \qquad \textrm{and} \qquad Z_3=2Z_{31}-Z_{34}
\end{equation}
vanish at 
\begin{equation}\label{nA.15}
s_n=\frac{2n\pi}{q\sqrt{4-\beta}}=2n\pi\,\left(\frac{r_0^3}{M}\right)^{1/2}
\left(\frac{r_0-3M}{r_0-6M}\right)^{1/2}, \qquad n=1,2,\ldots,.
\end{equation}
Accordingly, besides the sequence (50) found previously, there is a second infinite sequence of 
points $Q'_n(s_n)$ conjugate to $P_0$ on the geodesic B. To check whether the first of these, $Q'_1(s_1)$, is located on the segment $P_0P_1$ we compare the value of $s_1$ with the length 
$s_B$ of the curve from $P_0$ to $P_1$. From (8) one gets 
\begin{equation}\label{nA.16}
\frac{s_1}{s_B}=
\left(\frac{r_0}{r_0-6M}\right)^{1/2}>1
\end{equation}
showing that all the points $Q'_n$ lie beyond the relevant segment. One concludes that these 
conjugate points do not account for the fact that the circular geodesic B has a non-maximal 
length.

\end{document}